\documentclass[twocolumn,  
showpacs,  
preprintnumbers,  
aps,  
prd,  
a4paper,  
superscriptaddress,  
nofootinbib,  
tightenlines,  
floats,floatfix  
]{revtex4}

\usepackage{amsmath}
\usepackage{amsfonts}
\usepackage{graphicx}
\newcommand{\bea}{\begin{eqnarray}}
\newcommand{\ena}{\end{eqnarray}}
\begin{document}
\begin{titlepage}

\preprint{KU-TP 012}
\preprint{arXiv:astro-ph/0702015v3}

\begin{center}
\vspace*{5mm} {\Large \bf Probing the Coupling between Dark
Components of the Universe}\\

\vspace{10mm}
{\large\bf Zong-Kuan Guo,$^{a,}$\footnote{e-mail address: guozk@phys.kindai.ac.jp}
Nobuyoshi Ohta$^{a,}$\footnote{e-mail address: ohtan@phys.kindai.ac.jp}
and Shinji Tsujikawa$^{b,}$\footnote{e-mail address: shinji@nat.gunma-ct.ac.jp}
}\\
\vspace{5mm}
{\it
$^a$ Department of Physics, Kinki University, Higashi-Osaka,
 Osaka 577-8502, Japan \\
$^b$ Department of Physics, Gunma National College of
Technology, Gunma 371-8530, Japan}
\end{center}

\vspace{5mm}
\centerline{\large \bf Abstract}

\vspace{5mm}

We place observational constraints on a coupling between dark
energy and dark matter by using 71 Type Ia supernovae (SNe Ia)
from the first year of the five-year Supernova Legacy Survey
(SNLS), the cosmic microwave background (CMB) shift parameter from
the three-year Wilkinson Microwave Anisotropy Probe (WMAP), and
the baryon acoustic oscillation (BAO) peak found in the Sloan
Digital Sky Survey (SDSS). The interactions we study are (i)
constant coupling $\delta$ and (ii) varying coupling $\delta(z)$
that depends on a redshift $z$, both of which have simple
parametrizations of the Hubble parameter to confront with
observational data. We find that the combination of the three
databases marginalized over a present dark energy density gives
stringent constraints on the coupling, $-0.08 < \delta < 0.03$
(95\% CL) in the constant coupling model and $ -0.4 < \delta_0 <
0.1$ (95\% CL) in the varying coupling model, where $\delta_0$ is
a present value. The uncoupled $\Lambda$CDM model ($w_X=-1$ and
$\delta=0$) still remains a good fit to the data, but the negative
coupling ($\delta<0$) with the equation of state of dark energy $w_X<-1$
is slightly favoured over the $\Lambda$CDM model.

\vspace{2mm}
\begin{flushleft}
PACS number(s): 98.80.Es, 98.80.Cq
\end{flushleft}
\end{titlepage}

\vspace{5mm}

\section{Introduction}

Recent observations of Supernova Ia (SNIa) suggest that the
universe has entered a stage of an accelerated expansion
with a redshift $z \lesssim 1$ \cite{rie98, per99,ast05}.
This has been confirmed by precise measurements of the spectrum
of the Cosmic Microwave Background (CMB) anisotropies \cite{spe03,spe06}
as well as the baryon acoustic oscillations (BAO)
in the Sloan Digital Sky Survey (SDSS) luminous galaxy
sample \cite{eis05}. As is well known, all usual types of
matter with positive pressure generate attractive forces, which
decelerate the expansion of the universe. Given this, a dark
energy component with negative pressure was suggested to account
for the invisible fuel that drives the current accelerated
expansion (see Refs.~\cite{review,CST} for reviews).

The simplest candidate for dark energy is a cosmological constant
$\Lambda$ (vacuum energy), which corresponds to a constant equation of
state $w_X=-1$. This model, the so-called $\Lambda$CDM, provides
an excellent fit to a wide range of astronomical data so far.
However, in such a model there exists a theoretical problem of cosmic
coincidence: why is the vacuum density comparable with the
critical density at the present epoch in the long history of the
universe? One possible approach to alleviating this problem is to
assume that the ``cosmological constant" is not a constant but
is a dynamical component with a slowly evolving
and spatially homogeneous scalar field called quintessence \cite{quin}
(see Refs.~\cite{early} for early works and Refs.~\cite{rec} for
the reconstruction of quintessence potentials).
In such models the resolution of the cosmic coincidence problem
typically leads to a fine-tuning of model parameters.

Given the fact that the amount of dark matter is comparable to
that of dark energy in the present universe, it is natural to consider
an interaction between the two components. Originally
cosmological consequences of a scalar field coupled to the matter
were studied in Ref.~\cite{coupleearly}. Amendola \cite{ame00}
considered interaction between a quintessence field $\phi$ and
dark matter with a coupling $Q$ that satisfies the relation
$T^{\mu}_{\nu (\phi); \mu}=QT_{(m)}\phi_{;\mu}$ between energy
momentum tensors. In fact this type of interaction appears in
the context of scalar-tensor theories \cite{stensor},
$f(R)$ gravity models \cite{fR}, varying mass
dark matter/neutrino models \cite{Come,Huey,das05,neutrino}
and phantom dark energy models \cite{guo04}.
Interestingly the equation of state of dark energy can extend to
the region $w_X<-1$ if the mass of dark matter depends on
a quintessence field \cite{Huey,das05}.
The presence of the coupling can also provide an
accelerated scaling attractor along which the ratio of the energy
densities of dark energy and dark matter is a constant. Then this
may be useful to solve the coincident problem because the present
universe can be a global attractor with a dark energy fraction
$\Omega_X \simeq 0.7$.

However, the coupling $Q$ to realize the accelerated scaling attractor
is too large to ensure the presence of a standard matter dominated
epoch \cite{ame00}. In fact there exists the so-called ``$\phi$
matter-dominated epoch'' ($\phi$MDE) during which an effective equation
of state is given by $w_{\rm eff}=2Q^2/3$ \cite{ame00,CST,GNST}.
Amendola \cite{ame00} showed that the coupled quintessence model
with an exponential potential is not consistent with observational data of
CMB unless the coupling $Q$ is smaller than the order of 0.1,
but in this case there is no scaling accelerated attractor.
Instead the system finally approaches a scalar-field dominated attractor
with $w_{\rm eff}=-1$ and $\Omega_X=1$, in which case the
coincident problem is not solved.

There are many other scalar-field dark energy models such as K-essence,
tachyon, phantom and dilatonic ghost condensates. For a general Lagrangian
density $p(\phi, X)$ with a kinetic term $X=-(1/2)(\nabla \phi)^2$
and a constant coupling $Q$, the existence of scaling solutions
required to solve the coincident problem restricts the Lagrangian
density to the form $p=X\,g(Xe^{\lambda \phi})$, where $g$ is
an arbitrary function and $\lambda$ is a constant \cite{Piazza}.
Recently it has been shown that for the vast class of this generalized
Lagrangian that include most of scalar-field dark energy models,
the matter era is not followed by the accelerated scaling attractor~\cite{AQTW}.
Thus it is still a challenging task to construct coupled scalar-field models
which can solve the coincidence problem without using fine-tuned varying
couplings \cite{stationary}.

Since the origin of dark energy is not yet known, there are
several different approaches \cite{dal01,zim01,maj04,Wei,ame06} to
implementing couplings without restricting to scalar-field models
(see also Refs.~\cite{Pavo}-\cite{man05} for a number of
interesting aspects of interacting dark energy). The approach we
adopt in this paper is to introduce an interaction of the form
$\Gamma \rho_m$ on the rhs of conservation equations, see
(\ref{be1}) and (\ref{be2}). While this is basically a fluid
description of dark energy, Eqs.~(\ref{be1}) and (\ref{be2})
include the aforementioned scalar-field coupling by setting
$\Gamma=Q\dot{\phi}$. Since the interaction rate $\Gamma$ measured
by the Hubble rate $H$ is generally important to discuss the
strength of an energy transfer, we introduce a dimensionless
coupling $\delta$ in the form $\delta=\Gamma/H$. This is an
approach a number of authors adopted \cite{dal01}. One can
constrain the strength of the interaction observationally by
assuming that $\delta$ is a constant (as in the constant $Q$ case
discussed above). In fact the authors in Ref.~\cite{ame06}
recently placed observational constraints on the coupling by using
the SNIa data with a parametrization of the dark energy equation
of state $w_{X}=w_0+w_1z$.

It is also possible to address the varying $\delta$ case.
For example, Dalal {\it et al.} \cite{dal01} assumed that the ratio of
dark energy and dark matter has a relation $\rho_X/\rho_m \propto a^{\xi}$,
where $a$ is a scale factor and $\xi$ is a constant.
For a constant equation of state of dark energy, the coupling
$\delta$ is known as a function of the redshift $z$ \cite{maj04},
see Eq.~(\ref{del}). Since the strength of the coupling
decreases for larger $z$, this scenario can address the
situation in which the interaction is weak during the
matter era but becomes strong in the dark energy dominated epoch.

Observational constraints on the coupling $\delta=\Gamma/H$ have
been obtained by using SNIa data \cite{maj04,ame06}.
In this paper, we carry out likelihood analysis of coupled
dark energy models by using 71 high-redshift SNe Ia from the first
year of the five-year SNLS, the CMB shift parameter from the
three-year WMAP observations and the BAO peak found in the SDSS.
We concentrate on two classes of interacting models: (i) a
constant coupling $\delta$ and (ii) a varying coupling $\delta(z)$
with the relation $\rho_X/\rho_m \propto a^{\xi}$. Throughout this
paper, the dark energy equation of state $w_X$ is assumed to be a
constant. In both models, we have three free parameters $(\delta,
w_X, \Omega_{X0})$, where $\Omega_{X0}$ is the present energy
fraction of dark energy (in the varying coupling model $\delta$ is
replaced by the present value $\delta_0$). We find that the
combination of the three databases places stringent constraints on
the model parameters since the CMB shift parameter is sensitive to
the value of the coupling. Our results also indicate that the
concordance $\Lambda$CDM model still remains a good fit to the
data, but the negative coupling ($\delta<0$) with $w_X<-1$
is slightly favoured over the $\Lambda$CDM model.

\section{Interactions between dark energy and dark matter}

In this section, we explain the form of the interaction between
dark energy and dark matter.
The background metric is described by the flat
Friedmann-Robertson-Walker (FRW) metric
with a scale factor $a$:
\begin{eqnarray}
{\rm d}s^2=-{\rm d}t^2+a^2(t){\rm d}{\bf x}^2\,,
\end{eqnarray}
where $t$ is a cosmic time.
Quite generally we can write the conservation
equations in the forms
\begin{eqnarray}
\label{be1}
& &\dot{\rho}_m+3H \rho_m=+\Gamma \rho_m\,, \\
\label{be2}
& &\dot{\rho}_X+3H(\rho_X+p_X)=-\Gamma \rho_m\,,
\end{eqnarray}
where $H=\dot{a}/a$ is a Hubble rate,
$\rho_m$ and $\rho_X$ are the energy densities of dark
matter and dark energy respectively, and $p_X$ is the dark energy
pressure density with the equation of state $w_X=p_X/\rho_X$. If we
consider a scalar-field model of dark energy, the interaction term
is typically given by $\Gamma=Q\dot{\phi}$, where the constant $Q$
characterizes the strength of the coupling \cite{ame00,GNST}.
Amendola obtained the constraint on the coupling as $Q<0.08$
by using the information of the CMB power spectrum.

We note that the origin of dark energy is not yet identified
as a scalar field.
In this work we take a different approach to constraining
the strength of the interaction without assuming scalar-field models.
We measure $\Gamma$ in terms of the Hubble parameter
$H$ and define the dimensionless coupling
\begin{eqnarray}
\delta=\Gamma/H\,.
 \label{measure}
\end{eqnarray}
Note that a positive $\delta$ implies a
transfer of energy from dark energy to dark
matter, and vice versa.
{}From Eqs.~(\ref{be1}) and (\ref{be2}) it is clear that
the total energy density is conserved.
Neglecting the contributions of (uncoupled) baryon
and radiation components
the Friedmann equation is given by
\begin{eqnarray}
3H^2=\kappa^2 (\rho_m+\rho_X)\,,
\label{fe}
\end{eqnarray}
where $\kappa^2 = 8\pi G$ with $G$ being gravitational constant.

In the following, we first discuss the case in which $\delta$ is a
constant. Then the analysis is extended to the case in which
$\delta$ varies in time and the ratio of the energy densities of
dark energy and dark matter scales as
$\rho_X/\rho_m \propto a^{\xi}$.

\subsection{Constant coupling models}

For constant $\delta$~\cite{Wei,ame06}, Eq.~(\ref{be1}) is
easily integrated to give
\begin{eqnarray}
\label{rhom}
\rho_m=\rho_{m0}a^{-3+\delta}
=\rho_{m0} (1+z)^{3-\delta}\,,
\end{eqnarray}
where the subscript ``0'' represents the present values. Note that
$z$ is a redshift which is defined by $z = a_0/a-1$, where $a_0$ is
normalized as $a_0=1$.
Equation (\ref{rhom}) shows that the interaction leads to a deviation
from the usual conservation relation $\rho_m \propto a^{-3}$.

We assume that $w_X$ is a constant.
Then substituting Eq.~(\ref{rhom}) into Eq.~(\ref{be2}),
we obtain the following integrated solution
\begin{eqnarray}
\label{rhoXc}
\rho_X &=&\rho_{X0} (1+z)^{3(1+w_X)} \nonumber \\
& & +\rho_{m0} \frac{\delta}{\delta+3w_X}
\left[ (1+z)^{3(1+w_X)}-(1+z)^{3-\delta}\right].
\nonumber \\
\end{eqnarray}
Using the Friedmann equation (\ref{fe}), we find
\begin{eqnarray}
E^2(z) &=& \Omega_{X0}(1+z)^{3(1+w_X)} \nonumber \\
&&\hspace{-5mm}+\frac{1-\Omega_{X0}}{\delta+3w_X} \left[ \delta
(1+z)^{3(1+w_X)}+3w_X (1+z)^{3-\delta} \right],
\nonumber \\
 \label{model1}
\end{eqnarray}
where $E(z) = H(z)/H_0$ and $\Omega_{X0} =\kappa^2
\rho_{X0}/(3H_0^2)$. Thus we have three free parameters $(\delta,
w_X, \Omega_{X0})$ when we confront models with observations.
This allows us to parameterize a wide range of possible
cosmologies in a simple fashion.

In the high redshift region ($z \gg 1$), it follows from Eq.~(\ref{rhoXc})
that $\rho_X$ behaves as $\rho_X \simeq -\rho_{m0}\,\delta/(\delta+3w_X)
(1+z)^{3-\delta}$ for $3w_X<-\delta$. This means that the energy
density of dark energy becomes negative for $\delta<0$.
Since such a negative energy appears in phantom models \cite{phantom}
and also modified gravity models \cite{fR,APT2},
we do not exclude the possibility of the negative coupling.
Note also that in the high redshift region we have the scaling
relation $\rho_X/\rho_m=-\delta/(\delta+3w_X)$.
If $|\delta|$ is much smaller than 1, the ratio satisfies the
relation $|\rho_X/\rho_m| \ll 1$
provided that $w_X$ is of order $-1$.

\begin{figure}
\begin{center}
\includegraphics[width=70mm]{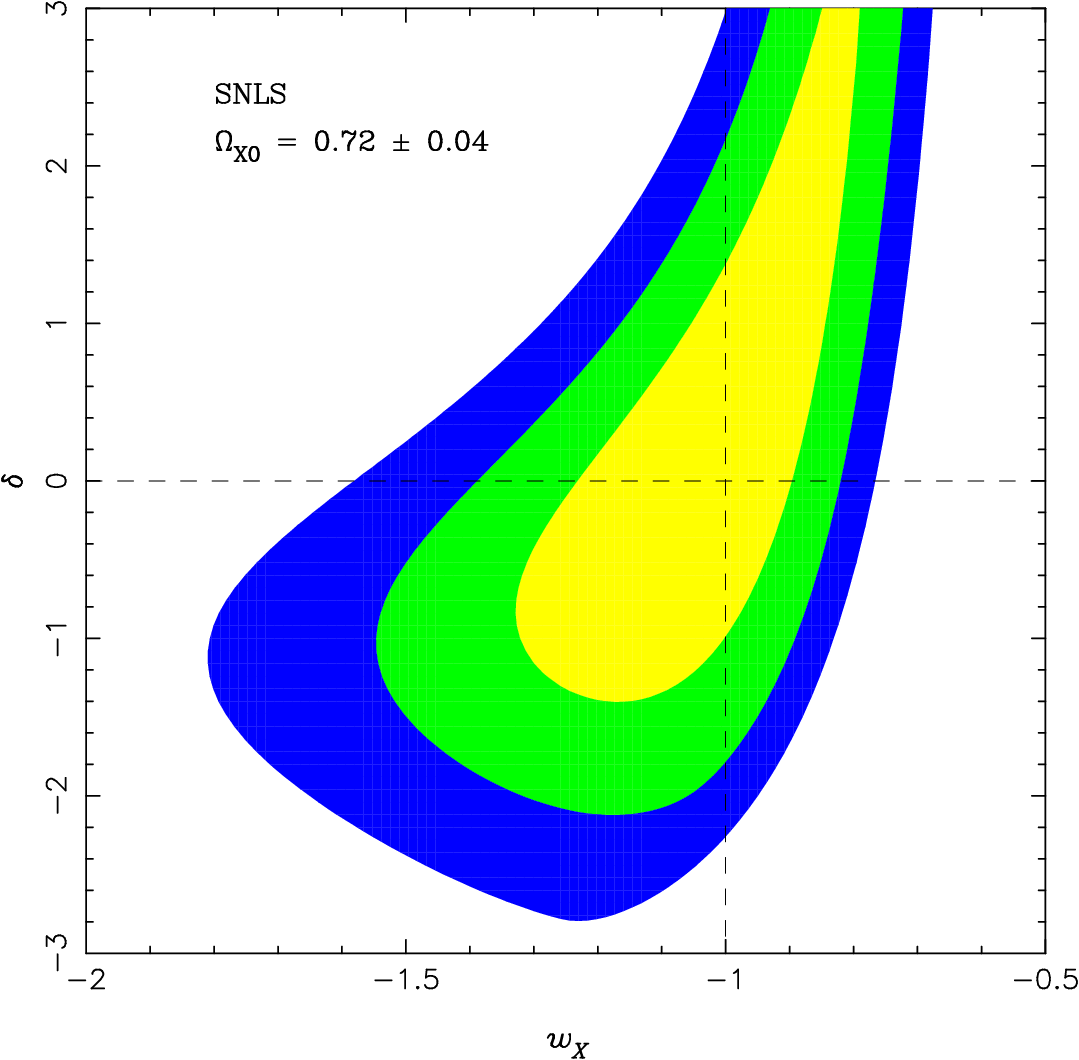}
\caption{Probability contours from the SNLS data only
at 68.3\%, 95.4\% and 99.7\%
confidence levels in the ($w_X, \delta$) plane
marginalized over $\Omega_{X0}$ with priors
$\Omega_{X0}=0.72 \pm 0.04$ and $\delta<3$
in the constant coupling models.
The horizontal and dashed lines
represent the uncoupled ``XCDM'' models and the coupled
$\Lambda$CDM models respectively, and their crossing point
corresponds to the standard $\Lambda$CDM model.
In this case we have the constraint $-1.78 <\delta<3$ (95\% CL). } \label{consn}
\end{center}
\end{figure}

With the parametrization given above, we can classify the models into the
following four types in the ($w_X$,$\delta$) plane: (i) decaying
phantom characterized by $\delta>0$ and $w_X<-1$, (ii) decaying
quintessence\footnote{Here we use the word ``quintessence'' for dark energy
satisfying $w_X>-1$ without restricting to scalar-field models.}
characterized by $\delta>0$ and $w_X>-1$, (iii)
created quintessence characterized by $\delta<0$ and $w_X>-1$,
and (iv) created phantom characterized by $\delta<0$ and
$w_X<-1$, as shown in Fig.~\ref{consn}.
In this plane, the horizontal dashed line represents
the uncoupled models with $\delta=0$ and
the vertical dashed line represents the coupled
$\Lambda$CDM models with $w_X=-1$.

\subsection{Varying coupling models}

When $\delta$ varies in time, Eq.~(\ref{be1}) together with
Eq.~(\ref{measure}) gives
\bea
\rho_m = \rho_{m0} a^{-3}
e^{\int
\delta\; {\rm d}(\ln a)}\,.
\ena
Let us now consider a situation in which the ratio of dark
energy and dark matter has the following relation \cite{dal01}:
\begin{equation}
\label{ratio}
\frac{\rho_X}{\rho_m} =
\frac{\rho_{X0}}{\rho_{m0}} a^{\xi},
\end{equation}
where $\xi$ is a constant which quantifies the severity of the
coincidence problem.
In the absence of the coupling $\delta$ with constant $w_X$,
the energy density of dark energy scales as
$\rho_X \propto a^{-3(1+w_X)}$.
Here the ratio $\rho_X/\rho_m$ is proportional to
$a^{-3w_X}$, namely,
the $\xi=-3w_X$ case in Eq.~(\ref{ratio}).
Note that the standard $\Lambda$CDM model
corresponds to $w_X=-1$ and $\xi=3$.

The general case $\xi \ne -3w_X$ indicates the
existence of an interaction between dark matter and dark energy.
Using the relation (\ref{ratio}), we find that the coupling $\Gamma$
is given by $\Gamma=-H(\xi+3w_X)\Omega_X(z)$, where
$\Omega_X(z)=\kappa^2 \rho_X/(3H^2)$.
This shows that $\delta$ varies according to the change
of $\Omega_X$ as
\begin{equation}
\delta (z)=-(\xi+3w_X) \Omega_X(z)\,.
\end{equation}
When $\xi=-3w_X$, this reduces to $\delta=0$.
Since $\Omega_X(z)$ is given by
$\Omega_X(z)=\left[ (\rho_{m0}/\rho_{X0})
(1+z)^\xi+1 \right]^{-1}$ under the condition (\ref{ratio}),
the coupling can be written as
\begin{equation}
\label{del}
\delta(z) = \frac{\delta_0}
 {\Omega_{X0} + (1-\Omega_{X0})(1+z)^\xi}\,,
\end{equation}
where $\delta_0 = - (\xi+3w_X)\Omega_{X0}$.
If $\xi>0$, $\delta(z)$ decreases for higher $z$.

Note that the $\xi=0$ case gives a constant coupling
$\delta (z)=\delta_0$. Since this corresponds to an exact
scaling solution $\rho_X \propto \rho_m$,
one cannot realize the matter dominated epoch followed
by a late-time acceleration. This constant $\delta$ case is
different from the one we discussed in subsection A.
In fact the solutions (\ref{rhom}) and (\ref{rhoXc})
do not satisfy the relation (\ref{ratio}).

{}From Eqs.~(\ref{be1}) and (\ref{be2}) together with
Eq.~(\ref{ratio}), the total energy density
$\rho_T=\rho_m+\rho_X$ satisfies
\begin{equation}
\frac{{\rm d} \ln \rho_T}{{\rm d} a} =  -\frac{3}{a}
 \left[1 + w_X \left(\frac{\rho_{m0}}{\rho_{X0}}a^{-\xi}
 + 1\right)^{-1}\right]\,,
\end{equation}
which can be integrated to give
\begin{equation}
\rho_T = \rho_{T0}\, a^{-3}
 \left[1-\frac{\rho_{X0}}{\rho_{T0}}
 (1-a^{\xi})\right]^{-3w_X/\xi}\,.
\end{equation}
Then the Friedmann equation~(\ref{fe}) gives
\begin{equation}
\label{model2}
 E^2(z) = (1+z)^3
 \left[1- \Omega_{X0} + \Omega_{X0}(1+z)^{-\xi}
 \right]^{-3w_X/\xi}\,.
\end{equation}
{}From Eq.~(\ref{ratio}), we find that the energy density
of dark energy is given by
\begin{equation}
\rho_X=\rho_T
\frac{\Omega_{X0}}{\Omega_{X0} + (1-\Omega_{X0})(1+z)^{\xi}}\,.
\end{equation}

We have three parameters ($\xi$, $w_X$, $\Omega_{X0}$)
in this model. Since $\delta_0$ is related to these variables
by the relation $\delta_0=-(\xi+3w_X)\Omega_{X0}$,
one can instead vary three parameters
($\delta_0$, $w_X$, $\Omega_{X0}$) when we carry
out likelihood analysis.
Thus the model (\ref{model2}) is a simple parametrization
that implements the variation of the coupling $\delta$.
Note that both $\rho_T$ and $\rho_X$ are positive
as long as $0<\Omega_{X0}<1$.
Hence $\rho_X$ remains positive in the high-redshift region
even for $\delta<0$ unlike the constant
coupling model.

The coupling affects the evolution of some quantities of interest,
such as the age of the universe and the deceleration parameter.
Given $w_X$ and $\Omega_{X0}$, the age of the
universe and the transition redshift at which the universe
switches from deceleration to acceleration become larger as the
value of $\delta_0$ increases from negative to positive.
In the next section, we place observational constraints
on the strength of the coupling.

\section{Constraints from Recent Observations}

In this section, we study the viability of interacting models presented in the
previous section by using recently released SNLS data~\cite{ast05}
in conjunction with the BAO peak in the SDSS~\cite{eis05}
and the CMB shift parameter~\cite{wan06}.

Recently Astier {\it et al.}~\cite{ast05} have compiled a new sample of
71 high-redshift SNe Ia, in the redshift range $0.2 < z < 1.0$,
discovered during the first year of the 5-year SNLS. This data set
is arguably the best high-redshift SN Ia compiled data, since the
multi-band, rolling search technique and careful calibration are
adopted. The luminosity distance $d_L(z)$ to supernovae
is given by \cite{review,CST}
\begin{equation}
d_L(z) = H_0^{-1}(1+z) \int_0^z \frac{{\rm d}z'}{E(z')}\,.
\end{equation}

The baryon oscillations in the galaxy power spectrum are imprints
from acoustic oscillations prior to recombination, which are also
responsible for the acoustic peaks seen in the CMB temperature
power spectrum. The physical length scale associated with the
oscillations is set by the sound horizon at recombination, which
can be estimated from the CMB data~\cite{spe06}. Measuring the
apparent size of the oscillations in a galaxy survey allows one to
measure the angular diameter distance at the survey redshift.
Although the acoustic features in the matter correlations are weak
on large scales, Eisenstein {\it et al.}~\cite{eis05} have
successfully found the peaks using a large spectroscopic sample of
luminous red galaxies from SDSS \cite{yor00}.
This sample contains 46,748 galaxies covering
3816 square degrees out to a redshift of $z=0.47$. They found a
parameter $A$, which is independent of dark energy
models~\cite{eis05}.
{}From Eq.~(5) in their paper \cite{eis05}, we write it as
\begin{equation}
A = \sqrt{\Omega_{m0}} \, E(z_1)^{-1/3}
 \left[\frac{1}{z_1}
 \int_0^{z_1}\frac{{\rm d}z'}{E(z')}\right]^{2/3},
\end{equation}
where $z_1 = 0.35$ and $A$ is measured to be $A = 0.469 \pm
0.017$. In our analysis, we combine these measurements.

The CMB shift parameter $\cal{R}$
captures the correspondence between the angular diameter
distance to last scattering surface and the relation of the
angular scale of the acoustic peaks to the physical scale
of the sound horizon \cite{bon97}.
Its value is expected to be mostly model-independent,
which can be extracted accurately from CMB data.
The shift parameter $\cal{R}$ is given by~\cite{wan04}
\begin{equation}
{\cal R} = \sqrt{\Omega_{m0}} \int_{0}^{z_{{\rm rec}}}
\frac{{\rm d}z'}{E(z')}\,,
\end{equation}
where $z_{\rm rec}$ is the redshift of recombination. It provides a
useful constraint on evolving dark energy models since the
integral over $E(z)$ extends to high redshifts.
The recent analysis of the three-year WMAP data \cite{spe06}
gives ${\cal R} = 1.70 \pm 0.03$
at $z_{\rm rec}=1089$ \cite{wan06}.

\subsection{Constant coupling models}

Let us first consider observational constraints on constant
coupling models. In Fig.~\ref{consn}, we show probability contours
from SNLS when we take a prior for $\Omega_{X0}$ such that the
probability distribution is a Gaussian with a mean of $0.72$ and a
standard deviation given by $\sigma=0.04$.
We also assume the condition $\delta<3$ to exclude the
possibility that dark matter behaves as a phantom matter [see Eq.~(\ref{rhom})].
The phantom dark matter is problematic for successful structure formation.
Then we find that the SNLS data gives a weak constraint on the coupling
as $-1.78 < \delta < 3$ at the 95\% confidence level.
When we choose the wider range of the
prior for $\Omega_{X0}$, the constraint becomes weaker.

\begin{figure}
\begin{center}
\includegraphics[width=70mm]{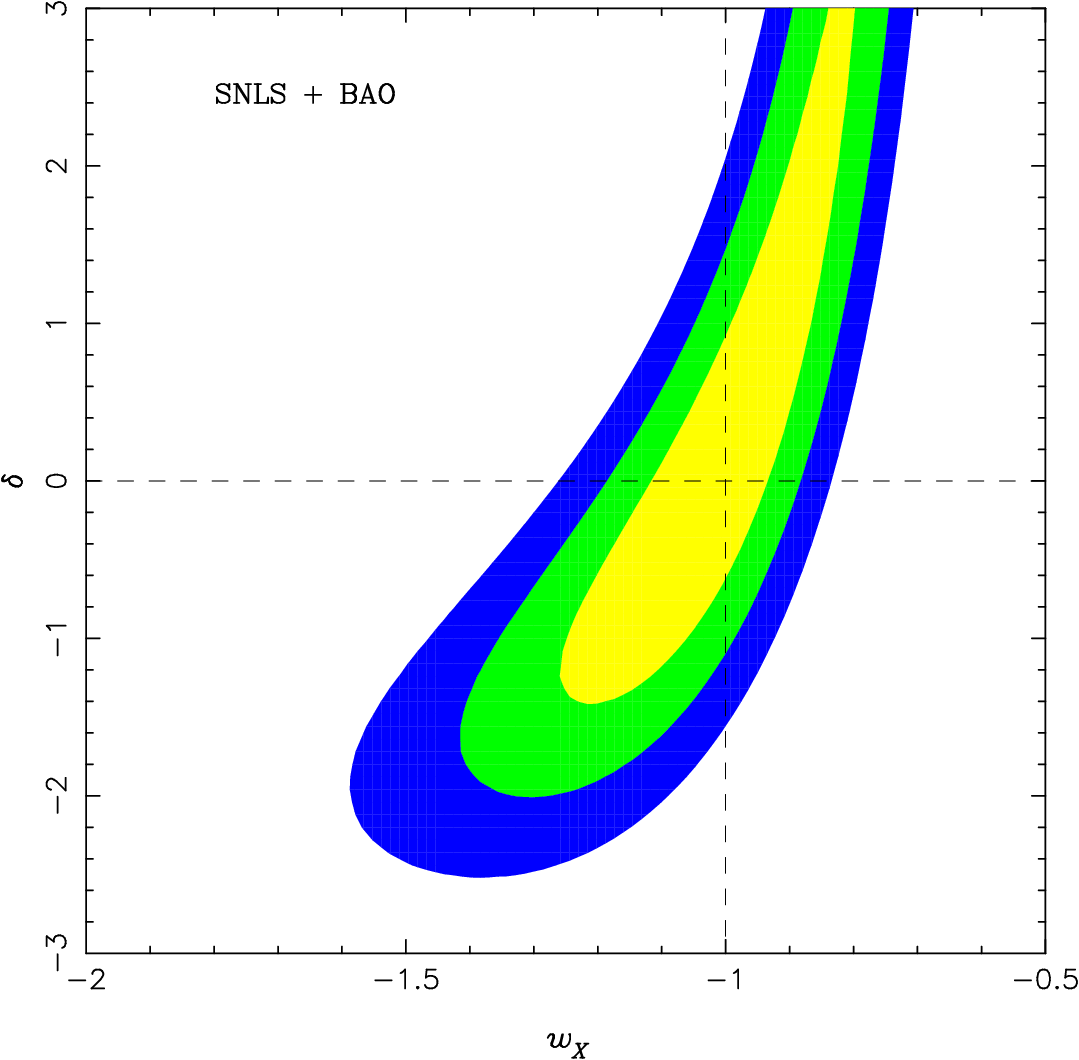}
\caption{Probability contours from the SNLS and
BAO data in the ($w_X, \delta$) plane marginalized over
$\Omega_{X0}$ without a prior for $\Omega_{X0}$
and with a prior $\delta<3$
in the constant coupling models.
In this case we have the constraint $-1.73 < \delta
<3$ (95\% CL). } \label{consnbao}
\end{center}
\end{figure}

Figure \ref{consnbao} shows the case in which the BAO data is
taken into account in addition to the SNLS data without a prior
for $\Omega_{X0}$. Compared to
Fig.~\ref{consn}, the allowed range of $w_X$ is reduced.
However we still have a rather large region of a parameter space for
$\delta$, i.e., $-1.73 < \delta < 3$ (95\% CL).

We have also carried out a likelihood analysis without
any prior for $\delta$ and found that even the large coupling
such as $\delta=20$ with $w_X \sim -0.7$ is within the
$2\sigma$ observational contour bound.
This reflects the fact that the SNIa and BAO data give
constraints only around low redshifts $z<{\cal O}(1)$.
The models can fit the data even for $\delta \gg 1$
because of the dominance of the $\delta (1+z)^{3(1+w_X)}$
term instead of the usual $3w_X (1+z)^{3-\delta}$ term
on the rhs of Eq.~(\ref{model1}).
This tells us how it is important to include other
data in a high-redshift region ($z \gg 1$) in order to rule out
models with problematic couplings ($\delta>3$).

\begin{figure*}
\begin{center}
\includegraphics[width=70mm]{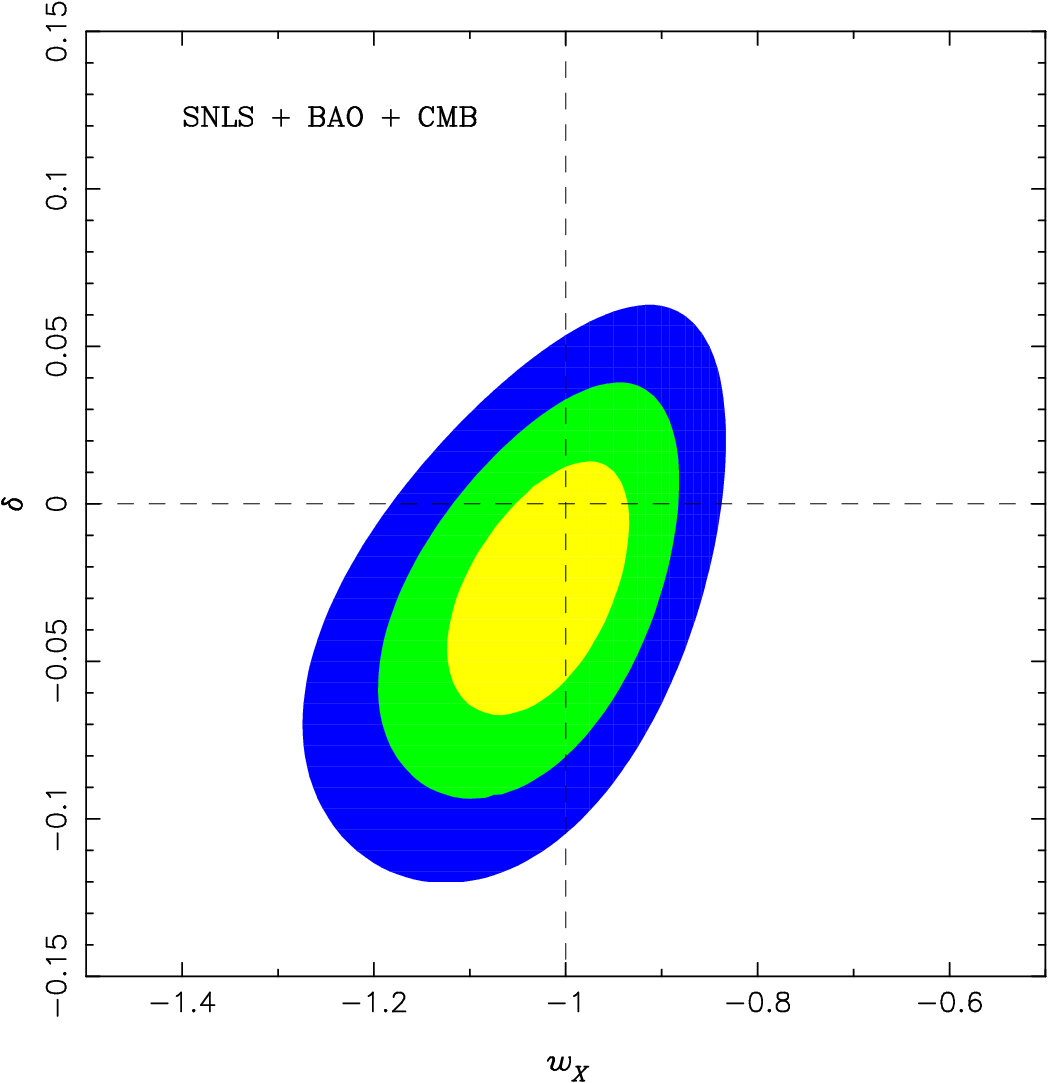}\hspace{5mm}
\includegraphics[width=70mm]{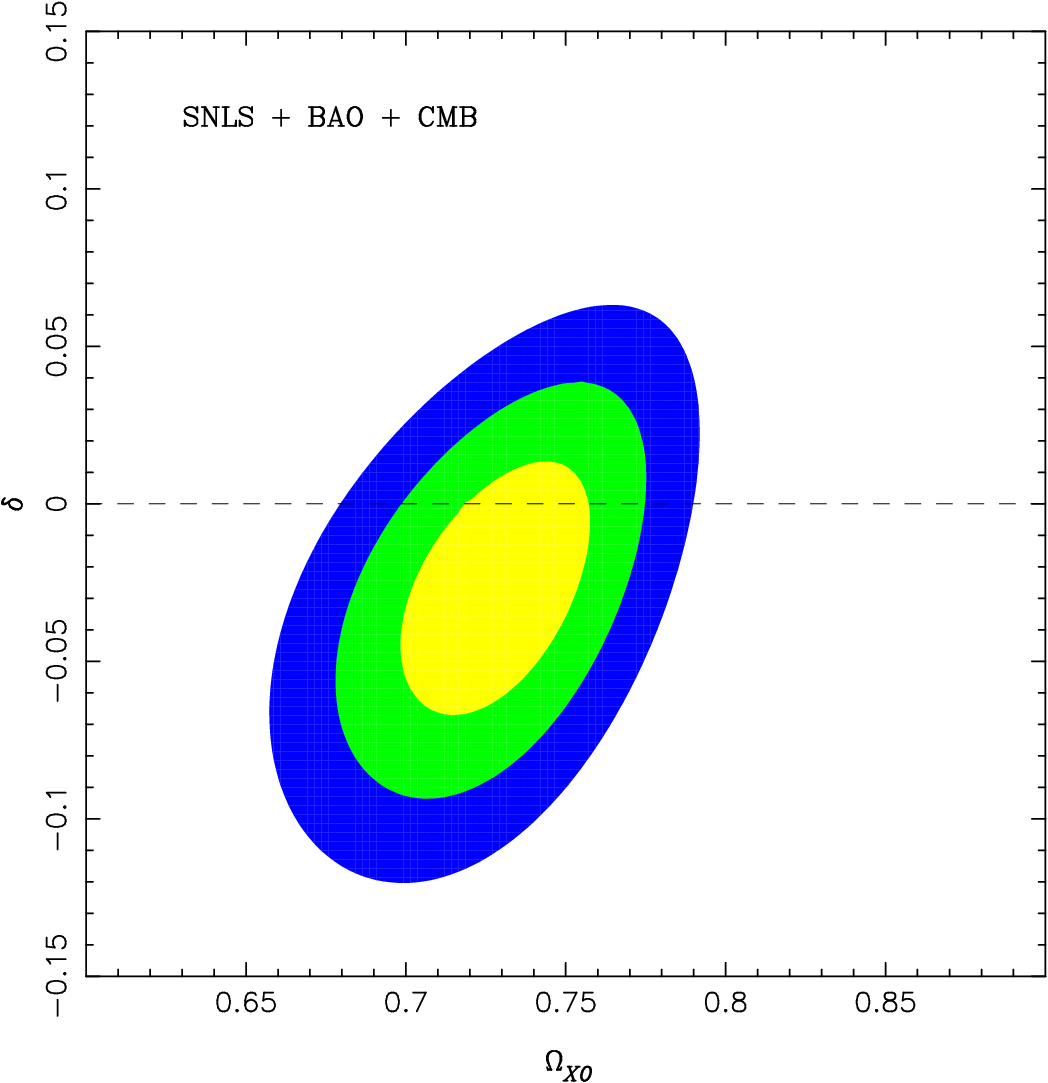}
\caption{Probability contours from the combination of
SNLS, BAO and CMB data in the constant coupling models.
The left panel shows observational contours
in the ($w_X, \delta$) plane marginalized over
$\Omega_{X0}$ without a prior for $\Omega_{X0}$,
whereas the right panel shows contours
in the ($\Omega_{X0}, \delta$)
plane marginalized over $w_X$ with no prior for $w_X$.
The best-fit model parameters correspond to
$\delta=-0.03$, $w_X=-1.02$ and
$\Omega_{X0}=0.73$ with $\chi^2=60.94$.
In this case we have the constraint
 $-0.08 < \delta < 0.03$ (95\% CL).}
\label{conall}
\end{center}
\end{figure*}

In fact the CMB shift parameter provides a stringent constraint
on the coupling. In Fig.~\ref{conall} we plot observational
contours from the joint analysis of SNLS, CMB and BAO data
in the ($w_X, \delta$) plane marginalized
over $\Omega_{X0}$ (left panel) and in the ($\Omega_{X0}, \delta$)
plane marginalized over $w_X$ (right panel).
Note that we do not put any prior for $\delta$ in these analysis.
We find that the coupling is severely constrained: $-0.08 < \delta < 0.03$
(95\% CL) in both marginalizations.
This comes from the fact that the CMB data do not allow
a large deviation from the standard matter-dominated epoch.


The combined analysis of three databases constrains the equation of
state and the present energy fraction of dark energy to $-1.16 < w_X < -0.91$
and $0.69 < \Omega_{X0} < 0.77$ (95\% CL).
It is interesting to note that the allowed observational contours are widely
spread in the phantom region ($w_X<-1$) with a negative coupling ($\delta<0$),
see Fig.~\ref{conall}.
The best-fit parameters are found to be $\delta=-0.03$, $w_X=-1.02$ and
$\Omega_{X0}=0.73$ with $\chi^2=60.94$,
which is slightly favoured over the $\Lambda$CDM model.

\subsection{Varying coupling models}

\begin{figure*}
\begin{center}
\includegraphics[width=70mm]{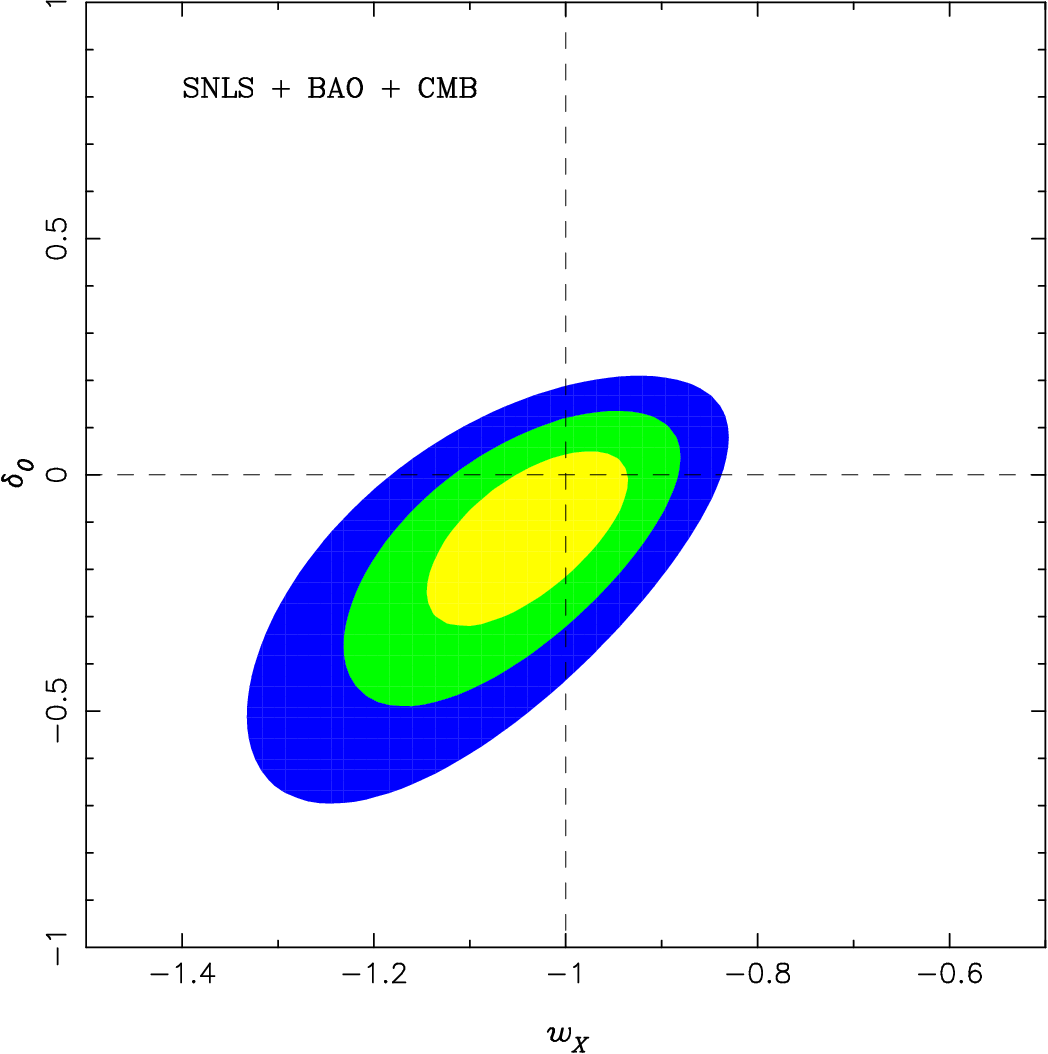}\hspace{5mm}
\includegraphics[width=70mm]{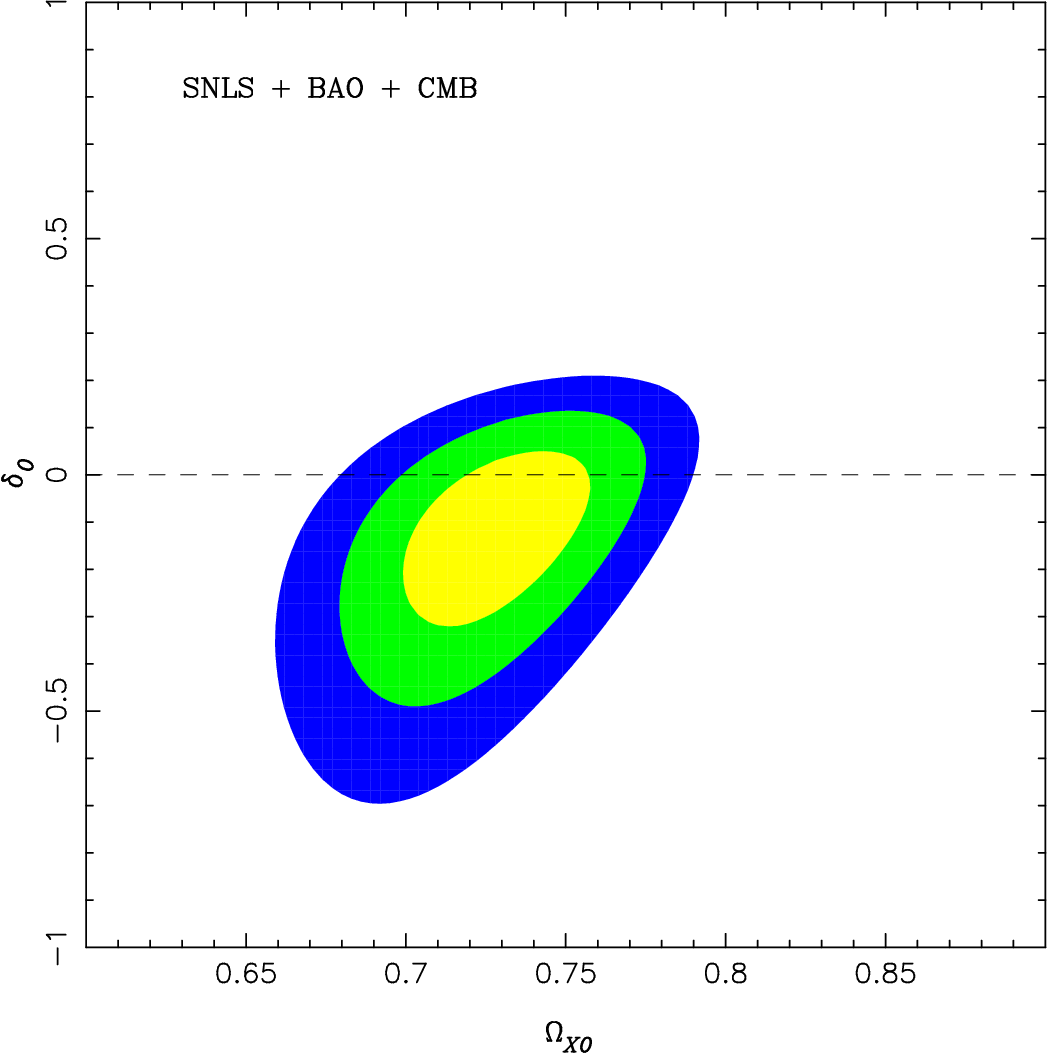}
\caption{Probability contours from the joint analysis of the
SNLS, BAO and CMB data in the varying coupling models.
The left panel shows observational contours in the
$(w_X, \delta_0)$ plane marginalized
over $\Omega_{X0}$ without prior for $\Omega_{X0}$,
whereas the right panel shows contours in the
($\Omega_{X0}, \delta$) plane marginalized over $w_X$
without prior for $w_X$.
The best-fit parameters correspond to $\delta=-0.11$, $w_X=-1.03$ and
$\Omega_{X0}=0.73$ with $\chi^2=60.94$.
In this case we have the constraint $-0.4 < \delta_0 < 0.1$ (95\% CL).}
\label{vaall}
\end{center}
\end{figure*}

We now proceed to the varying coupling models in which $\delta$
depends upon $z$ in the form (\ref{del}). In Fig.~\ref{vaall} we
show observational contours from the combined analysis of
SNLS$+$BAO$+$CMB data in the planes (i) $(w_X, \delta_0)$
marginalized over $\Omega_{X0}$ (left panel) and (ii) $(\Omega_{X0},
\delta_0)$ marginalized over $w_{X}$ (right panel).
We find that the present coupling, the dark energy equation of state
and the present dark energy density are constrained
to be $-0.4 < \delta_0 < 0.1$,
$-1.18 < w_X < -0.91$ and $0.69 < \Omega_{X0} < 0.77$ at the 95\%
confidence level. The best-fit parameters correspond to $\delta=-0.11$,
$w_X=-1.03$ and $\Omega_{X0}=0.73$ with $\chi^2=60.94$.
Similarly to the constant coupling case, the SNIa and BAO data
do not provide stringent constraints on $\delta_0$, but
inclusion of the CMB data significantly reduces the
allowed region of the coupling. Since $\delta(z)$ decreases for
larger $z$, the observational constraints on $\delta_0$ is not so
severe compared to the constant coupling models.
We also find that the phantom models ($w_X<-1$) with a negative
coupling ($\delta_0<0$) have a wider allowed parameter space
compared to three other divided regions
in the left panel of Fig.~\ref{vaall}.

\begin{figure}
\begin{center}
\includegraphics[width=70mm]{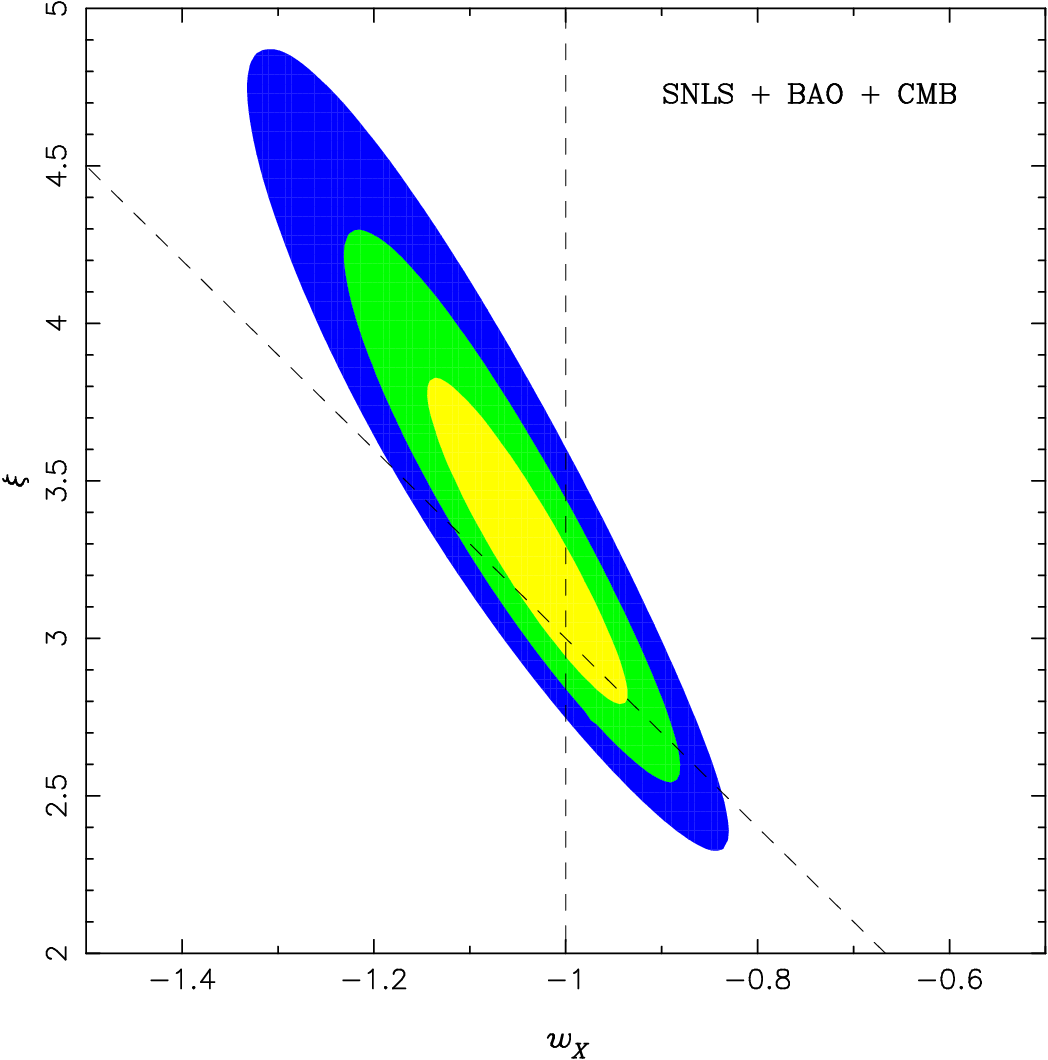}
\caption{Probability contours in the varying coupling models
in the $(w_X, \xi)$ plane
marginalized over $\Omega_{X0}$.
The line $\xi=-3w_X$ corresponds to the uncoupled models.
In this case we have the constraint
$2.66 < \xi < 4.05$ (95\% CL).}
\label{vaall2}
\end{center}
\end{figure}

As we see from Fig.~\ref{vaall2}, the $\Lambda$CDM model, which
corresponds to the point $(w_X,\xi)=(-1, 3)$, is within the
1$\sigma$ contour bound. We remind that the uncoupled models
are characterized by the line $\xi=-3w_X$. Thus, provided that the points
are not on the line $\xi=-3w_X$, the coupled models are allowed
observationally in the parameter regions $2.66 < \xi < 4.05$
(95\% CL). {}From Fig.~\ref{vaall2} it is obvious that the scaling
models with $\xi=0$ are excluded from the data.

\section{Conclusions}

In this paper, we have studied interacting models of dark energy
in which the coupling $\delta$ between dark matter and dark energy
is given by (\ref{measure}) together with conservation equations
(\ref{be1}) and (\ref{be2}).
We discussed two different models: (i) constant
$\delta$ case, and (ii) varying $\delta$ case which has
a redshift dependence given in (\ref{del}).
The latter type of couplings appears by imposing the
relation $\rho_X/\rho_m \propto a^\xi$ between
the energy densities of two dark components.
Assuming that the equation of state of dark energy $w_X$
is a constant, we obtain the convenient forms of the Friedmann
equations (\ref{model1}) and (\ref{model2}) to confront with
observational data.

We have placed observational constraints on the strength of the
coupling $\delta$ by using the recent data of SNLS, the CMB shift
parameter and the SDSS baryon acoustic oscillations (BAO).
In both constant and varying coupling models, the supernova data
alone do not provide stringent constraints on the coupling.
Adding the BAO data generally reduces the allowed range of
$w_X$, but the coupling $\delta$ of order unity is not still
ruled out.  This reflects the fact that  the
BAO and SN Ia data are sensitive to the
value of $w_X$ or $\Omega_{X0}$ rather than $\delta$.
However the combined analysis of SNLS$+$BAO$+$CMB shows
that the allowed region of the coupling is significantly reduced
compared to the case without the CMB data.
This is associated with the fact that a large coupling leads to the
change of the cosmological evolution during the matter-dominated
epoch, thus modifying the CMB angular-diameter distance.

By the joint analysis of SNLS$+$BAO$+$CMB,
we have obtained observational constraints on
the strength of the coupling: (i) $-0.08 < \delta < 0.03$ (95\%
CL) for the constant coupling models and (ii) $-0.4 < \delta_0 <
0.1$ (95\% CL) for the varying coupling models (here $\delta_0$ is
a present value). We also find that the
best-fit values exist in the phantom region ($w_X < -1$) with a
negative coupling ($\delta <0$)
in both constant and varying coupling models.
The uncoupled $\Lambda$CDM model ($w_X=-1$ and $\delta=0$) still
remains a good fit to the data. Nevertheless it is interesting to
note that the negative coupling with a dark energy with $w_X<-1$
is slightly favoured over the $\Lambda$CDM model.

There exist coupled dark energy models \cite{Huey,das05}
that give $w_X<-1$ without using a negative kinetic energy
of a scalar field (called ``super-acceleration'' in Ref.~\cite{das05}).
This can be realized by considering scalar-field
dependent masses of dark matter particles.
It was shown in Ref.~\cite{das05} that this super-acceleration model
satisfies a number of observational constraints, which
is consistent with our results that the equation of state $w_X<-1$ is favoured.

In this work we did not take into account observational constraints
from matter density perturbations $\delta_m$ which is also affected by the
presence of the coupling \cite{mper}.
It was shown in Ref.~\cite{Fabris} that the inclusion of matter density
perturbations can place a stronger constraint on the coupling $\delta$
compared to the analysis using the background evolution only.
The future galaxy surveys such as
KAOS and PANSTARS will further provide good data of the matter
power spectrum with a high accuracy, which can offer a possibility
to put severe constraints on the coupling.
This may provide us an exciting possibility to reveal
the origin of dark energy and dark matter.

\section*{ACKNOWLEDGEMENTS}
We thank Luca Amendola for fruitful discussions.
We are also grateful to Orfeu Bertolami and Joan Sola
for useful correspondence.
The work of N.\,O. and Z.K.\,G. was supported in part by the Grant-in-Aid for
Scientific Research Fund of the JSPS Nos. 16540250 and 06042.
S.\,T. is supported by JSPS (Grant No.\,30318802).


\begin{thebibliography}{99}
\bibitem{rie98}
A.~G.~Riess {\it et al.},
Astron.\ J.\  {\bf 116}, 1009 (1998);
Astron.\ J.\  {\bf 117}, 707 (1999).

\bibitem{per99}
S.~Perlmutter {\it et al.},
Astrophys.\ J.\  {\bf 517}, 565 (1999).

\bibitem{ast05}
P.~Astier {\it et al.},
Astron.\ Astrophys.\  {\bf 447}, 31 (2006)

\bibitem{spe03}
D.~N.~Spergel {\it et al.}  [WMAP Collaboration],
Astrophys.\ J.\ Suppl.\  {\bf 148}, 175 (2003).

\bibitem{spe06}
D.~N.~Spergel {\it et al.},
arXiv:astro-ph/0603449.

\bibitem{eis05}
D.~J.~Eisenstein {\it et al.}  [SDSS Collaboration],
Astrophys.\ J.\  {\bf 633}, 560 (2005).

\bibitem{review}
V.~Sahni and A.~A.~Starobinsky,
Int.\ J.\ Mod.\ Phys.\ D {\bf 9}, 373 (2000);
V.~Sahni,
Lect.\ Notes Phys.\  {\bf 653}, 141 (2004);
S.~M.~Carroll,
Living Rev.\ Rel.\  {\bf 4}, 1 (2001);
T.~Padmanabhan,
Phys.\ Rept.\  {\bf 380}, 235 (2003);
P.~J.~E.~Peebles and B.~Ratra,
Rev.\ Mod.\ Phys.\  {\bf 75}, 559 (2003).

\bibitem{CST}
E.~J.~Copeland, M.~Sami and S.~Tsujikawa,
Int.\ J.\ Mod.\ Phys.\ D {\bf 15}, 1753 (2006).

\bibitem{quin}
R.~R.~Caldwell, R.~Dave and P.~J.~Steinhardt,
Phys.\ Rev.\ Lett.\  {\bf 80}, 1582 (1998);
I.~Zlatev, L.~M.~Wang and P.~J.~Steinhardt,
Phys.\ Rev.\ Lett.\  {\bf 82}, 896 (1999).

\bibitem{early}
Y.~Fujii,
Phys.\ Rev.\ D {\bf 26}, 2580 (1982);
L.~H.~Ford,
Phys.\ Rev.\ D {\bf 35}, 2339 (1987);
C.~Wetterich, Nucl. \ Phys \ B. {\bf 302},
668 (1988);
B.~Ratra and J.~Peebles,
Phys. \ Rev \ D {\bf 37}, 321 (1988);
Y.~Fujii and T.~Nishioka,
Phys.\ Rev.\ D {\bf 42}, 361 (1990).

\bibitem{rec}
A.~A.~Starobinsky,
JETP Lett.\  {\bf 68}, 757 (1998);
D.~Huterer and M.~S.~Turner,
Phys.\ Rev.\ D {\bf 60}, 081301 (1999);
T.~Chiba and T.~Nakamura,
Phys.\ Rev.\ D {\bf 62}, 121301 (2000);
M.~Sahlen, A.~R.~Liddle and D.~Parkinson,
Phys.\ Rev.\ D {\bf 72}, 083511 (2005);
Z.~K.~Guo, N.~Ohta and Y.~Z.~Zhang,
Phys.\ Rev.\ D {\bf 72}, 023504 (2005);
S.~Tsujikawa,
Phys.\ Rev.\ D {\bf 72}, 083512 (2005);
Z.~K.~Guo, N.~Ohta and Y.~Z.~Zhang,
Mod.\ Phys.\ Lett.\  A {\bf 22}, 883 (2007).

\bibitem{coupleearly}
J.~Ellis, S.~Kalara, K.~A.~Olive and C.~Wetterich,
Phys.\ Lett.\ B {\bf 228}, 264 (1989);
T.~Damour and K.~Nordtvedt, Phys.\ Rev.\ D {\bf 48}, 3436 (1993);
T.~Damour and A.~M.~Polyakov, Nucl. Phys. B {\bf 423}, 532 (1994).

\bibitem{ame00}
L. Amendola, Phys. Rev. {\bf D62}, 043511 (2000).

\bibitem{stensor}
L.~Amendola,
Phys.\ Rev.\ D {\bf 60}, 043501 (1999).

\bibitem{fR}
L.~Amendola, D.~Polarski and S.~Tsujikawa,
Phys.\ Rev.\ Lett.\  {\bf 98}, 131302 (2007).

\bibitem{Come}
M.~Doran and J.~Jaeckel,
Phys.\ Rev.\ D {\bf 66}, 043519 (2002);
D.~Comelli, M.~Pietroni and A.~Riotto,
Phys.\ Lett.\ B {\bf 571}, 115 (2003);
H.~Ziaeepour,
Phys.\ Rev.\ D {\bf 69}, 063512 (2004);
M.~Axenides and K.~Dimopoulos,
JCAP {\bf 0407} (2004) 010.

\bibitem{Huey}
G.~Huey and B.~D.~Wandelt,
Phys.\ Rev.\ D {\bf 74}, 023519 (2006).

\bibitem{das05}
S.~Das, P.~S.~Corasaniti and J.~Khoury,
Phys.\ Rev.\ D {\bf 73}, 083509 (2006).

\bibitem{neutrino}
P.~Q.~Hung, arXiv:hep-ph/0010126; M.~Li, X.~Wang, B.~Feng and
X.~Zhang, Phys.\ Rev.\ D {\bf 65}, 103511 (2002); M.~Li and
X.~Zhang, Phys.Lett. B {\bf 573}, 20 (2003); R.~Fardon,
A.~E.~Nelson and N.~Weiner, JCAP {\bf 0410}, 005 (2004); H.~Li,
B.~Feng, J.~Q.~Xia and X.~Zhang, Phys.\ Rev.\ D {\bf 73}, 103503
(2006);
A.~W.~Brookfield, C.~van de Bruck, D.~F.~Mota
and D.~Tocchini-Valentini,
Phys.\ Rev.\ Lett.\  {\bf 96}, 061301 (2006);
Phys.\ Rev.\ D {\bf 73}, 083515 (2006).

\bibitem{guo04}
Z.~K.~Guo and Y.~Z.~Zhang, Phys.\ Rev.\ D {\bf 71}, 023501 (2005);
R.~G.~Cai and A.~Wang,
JCAP {\bf 0503}, 002 (2005);
Z.~K.~Guo, R.~G.~Cai and Y.~Z.~Zhang, JCAP {\bf 0505}, 002 (2005);
R.~Curbelo, T.~Gonzalez and I.~Quiros,
Class.\ Quant.\ Grav.\ {\bf 23}, 1585 (2006);
B.~Chang {\it et al.}, JCAP {\bf 01}, 016 (2007).

\bibitem{GNST}
B.~Gumjudpai, T.~Naskar, M.~Sami and S.~Tsujikawa,
JCAP {\bf 0506}, 007 (2005).

\bibitem{Piazza}
F.~Piazza and S.~Tsujikawa,
JCAP {\bf 0407}, 004 (2004);
S.~Tsujikawa and M.~Sami,
Phys.\ Lett.\ B {\bf 603}, 113 (2004).

\bibitem{AQTW}
L.~Amendola, M.~Quartin, S.~Tsujikawa and I.~Waga,
Phys.\ Rev.\ D {\bf 74}, 023525 (2006).

\bibitem{stationary}
L.~Amendola and D.~Tocchini-Valentini,
Phys.\ Rev.\ D {\bf 64}, 043509 (2001).

\bibitem{dal01}
N.~Dalal, K.~Abazajian, E.~Jenkins and A.~V.~Manohar,
Phys.\ Rev.\ Lett.\  {\bf 87}, 141302 (2001).

\bibitem{zim01}
W.~Zimdahl, D.~Pavon and L.~P.~Chimento,
Phys. Lett. {\bf B521} (2001) 133;
W.~Zimdahl and D.~Pavon,
Gen.\ Rel.\ Grav.\  {\bf 35}, 413 (2003);
L.~P.~Chimento, A.~S.~Jakubi, D.~Pavon and W.~Zimdahl,
Phys.\ Rev.\ D {\bf 67}, 083513 (2003).

\bibitem{maj04}
E.~Majerotto, D.~Sapone and L.~Amendola,
arXiv:astro-ph/0410543.

\bibitem{Wei}
H.~Wei and S.~N.~Zhang,
Phys.\ Lett.\ B {\bf 644}, 7 (2007).

\bibitem{ame06}
L.~Amendola, G.~Camargo Campos and R.~Rosenfeld,
Phys.\ Rev.\  D {\bf 75}, 083506 (2007).

\bibitem{Pavo}
D.~Pavon, S.~Sen and W.~Zimdahl,
JCAP {\bf 0405}, 009 (2004);
G.~Olivares, F.~Atrio-Barandela and D.~Pavon,
Phys.\ Rev.\ D {\bf 71}, 063523 (2005);
Phys.\ Rev.\ D {\bf 74}, 043521 (2006);
S.~del Campo, R.~Herrera, G.~Olivares and D.~Pavon,
Phys.\ Rev.\ D {\bf 74}, 023501 (2006).


\bibitem{France}
U.~Franca and R.~Rosenfeld,
Phys.\ Rev.\ D {\bf 69}, 063517 (2004).

\bibitem{Maccio}
A.~V.~Maccio {\it et al.},
Phys.\ Rev.\ D {\bf 69}, 123516 (2004).

\bibitem{Nishi}
M.~Nishiyama, M.~a.~Morita and M.~Morikawa,
arXiv:astro-ph/0403571.

\bibitem{Noji}
S.~Nojiri, S.~D.~Odintsov and S.~Tsujikawa,
Phys.\ Rev.\ D {\bf 71}, 063004 (2005);
S.~Nojiri and S.~D.~Odintsov,
Phys.\ Rev.\ D {\bf 72}, 023003 (2005);
Gen.\ Rel.\ Grav.\  {\bf 38}, 1285 (2006).

\bibitem{Cal}
G.~Calcagni, S.~Tsujikawa and M.~Sami,
Class.\ Quant.\ Grav.\  {\bf 22}, 3977 (2005);
S.~Tsujikawa,
Phys.\ Rev.\ D {\bf 73}, 103504 (2006).

\bibitem{Das}
S.~Das and N.~Banerjee,
Gen.\ Rel.\ Grav.\  {\bf 38}, 785 (2006).

\bibitem{Cai}
H.~Wei and R.~G.~Cai, Phys.\ Rev.\ D {\bf 71}, 043504 (2005);
Phys.\ Rev.\ D {\bf 73}, 083002 (2006);
H.~Zhang and Z.~H.~Zhu, Phys.\ Rev.\ D {\bf 73} 043518 (2006);
H.~Wei and R.~G.~Cai, arXiv:astro-ph/0607064.

\bibitem{Xin}
X.~Zhang,
Phys.\ Lett.\ B {\bf 611}, 1 (2005);
Mod.\ Phys.\ Lett.\ A {\bf 20}, 2575 (2005).

\bibitem{Bin}
B.~Wang, Y.~g.~Gong and E.~Abdalla,
Phys.\ Lett.\ B {\bf 624}, 141 (2005);
D.~Pavon and W.~Zimdahl, Phys.\ Lett.\ B {\bf 628} 206 (2005);
B.~Wang, C.~Y.~Lin and E.~Abdalla,
Phys.\ Lett.\ B {\bf 637}, 357 (2006);
H.~Li, Z.~K.~Guo and Y.~Z.~Zhang, Int.\ J.\ Mod.\ Phys.\ D {\bf 15}, 869
(2006).

\bibitem{Koi}
T.~Koivisto,
Phys.\ Rev.\ D {\bf 72}, 043516 (2005).

\bibitem{Mota}
D.~F.~Mota and D.~J.~Shaw,
Phys.\ Rev.\ Lett.\  {\bf 97}, 151102 (2006).

\bibitem{Huang}
Z.~G.~Huang, H.~Q.~Lu and W.~Fang,
Class.\ Quant.\ Grav.\  {\bf 23}, 6215 (2006);
arXiv:hep-th/0610018.

\bibitem{Hu}
B.~Hu and Y.~Ling,
Phys.\ Rev.\ D {\bf 73}, 123510 (2006).

\bibitem{Lee}
S.~Lee, G.~C.~Liu and K.~W.~Ng,
Phys.\ Rev.\ D {\bf 73}, 083516 (2006).

\bibitem{Pop}
N.~J.~Poplawski,
Phys.\ Rev.\ D {\bf 74}, 084032 (2006).

\bibitem{Sad}
H.~M.~Sadjadi and M.~Alimohammadi,
Phys.\ Rev.\ D {\bf 74}, 103007 (2006);
H.~M.~Sadjadi and M.~Honardoost,
Phys.\ Lett.\  B {\bf 647}, 231 (2007).

\bibitem{Bono}
S.~A.~Bonometto, L.~Casarini, L.~P.~L.~Colombo and R.~Mainini,
arXiv:astro-ph/0612672.

\bibitem{Tet}
N.~Tetradis, J.~D.~Vergados and A.~Faessler,
Phys.\ Rev.\ D {\bf 75}, 023504 (2007).

\bibitem{Rosen}
R.~Rosenfeld,
Phys.\ Rev.\  D {\bf 75}, 083509 (2007).

\bibitem{Setare}
M.~R.~Setare,
 Phys.\ Lett.\ B {\bf 642}, 1 (2006);
 Phys.\ Lett.\ B {\bf 642}, 421 (2006);
 JCAP {\bf 01}, 023 (2007);
 arXiv:hep-th/0701085.

\bibitem{bea01}
R.~Bean and J.~Magueijo, Phys.\ Lett.\ B {\bf 517}, 177 (2001);
R.~Bean, Phys.\ Rev.\ D {\bf 64} 123516 (2001).

\bibitem{szy05}
 M.~Szydlowski, Phys.\ Lett.\ B {\bf 632}, 1 (2006);
 M.~Szydlowski, T.~Stachowiak and R.~Wojtak,
 Phys.\ Rev.\ D {\bf 73}, 063516 (2006).

\bibitem{man05}
 M.~Manera and D.~F.~Mota, Mon.\ Not.\ Roy.\
 Astron.\ Soc. {\bf
371}, 1373 (2006);
 A.~Arbey, Phys.\ Rev.\ D {\bf 74}, 043516 (2006);
 A.~Arbey, arXiv:astro-ph/0506732;
 M.~C.~Bento et al., Phys.\ Rev.\ D {\bf 73}, 103521 (2006);
 M.~C.~Bento et al., Phys.\ Rev.\ D {\bf 73}, 043504 (2006).


\bibitem{phantom}
R.~R.~Caldwell, Phys.\ Lett.\ B {\bf 545}, 23 (2002).

\bibitem{APT2}
L.~Amendola, R.~Gannouji, D.~Polarski and S.~Tsujikawa,
Phys.\ Rev.\  D {\bf 75}, 083504 (2007).

\bibitem{wan06}
Y.~Wang and P.~Mukherjee, Astrophys. J. {\bf 650}, 1 (2006).

\bibitem{yor00}
D.~G.~York et al., Astron. J. {\bf 120}, 1579 (2000).

\bibitem{bon97}
J.~R.~Bond, G.~Efstathiou and M.~Tegmark,
Mon.\ Not.\ Roy.\ Astron.\ Soc.\  {\bf 291}, L33 (1997).

\bibitem{wan04}
Y.~Wang and P.~Mukherjee, Astrophys. J. {\bf 606}, 654 (2004).

\bibitem{mper}
L.~Amendola,
Phys.\ Rev.\ D {\bf 69}, 103524 (2004);
L.~Amendola, S.~Tsujikawa and M.~Sami,
Phys.\ Lett.\ B {\bf 632}, 155 (2006).

\bibitem{Fabris}
J.~C.~Fabris, I.~L.~Shapiro and J.~Sola,
JCAP {\bf 0702}, 016 (2007).

\end{thebibliography}
\end{document}